\documentclass[9pt]{book}

\usepackage[dvips]{graphicx,color}
\usepackage{makeidx,universe}

\makeauthorindex

\BookTitle{The proceedings of the Physics of Accreting Compact Binaries}

\CopyRight{\copyright 2010 by Universal Academy Press, Inc.}

\begin{document} 

\pagenumbering{arabic}

\chapter{%
Accretion Disks in Evolved Cataclysmic Variables}

\author{\raggedright \baselineskip=10pt%
{\bf Sergey Zharikov,$^{1}$
Gaghik Tovmassian,$^{1}$
Andres Aviles$^{1}$, \\
Mauricio Tapia$^{1}$
and
Miguel Roth$^{2}$} \\
{\small \it %
(1) Instituto de Astronomia, Universidad Nacional Autonoma de Mexico, Ensenada, BC, Mexico \\
(2) Las Campanas Observatory, Carnegie Instututio of Washington, Casilla 601, La Serena, Chile 
}
}


\AuthorContents{Sergey Zharikov, Gaghik Tovmassian,  Andres Aviles, Maurisio Tapia and M. Roth} 

\AuthorIndex{Zharikov}{S.} 

\AuthorIndex{Tovmassian}{G} 

\AuthorIndex{Aviles}{A} 

\AuthorIndex{Tapia}{M}

\AuthorIndex{Roth}{M}

     \baselineskip=10pt
     \parindent=10pt

\section*{Abstract} 

We explore conditions and structure of accretion disks in short-period Cataclysmic Variables (CVs),
which have evolved beyond the period minimum. We show that accretion discs in systems with 
extreme mass ratios grow up to the size of corresponding Roche lobe and are relatively cool. 
In contrast, the viscosity and temperature in spiral arms formed as a result of a 2:1 resonance are
higher and their contribution plays an increasingly important role. We model such discs and generate light curves which  successfully simulate the observed double-humped light curves of 
SDSS1238, SDSS0804, SDSS1610 and V 455 And in quiescence.

\section{Introduction} 

 
 {According to  population models, there should be a significant number of systems near the period minimum  \cite{1999MNRAS.309.1034K}.
The  period distribution of  CVs, which include new systems discovered in the Sloan Digital Sky survey (SDSS), confirms the prediction 
about a accumulation of CV systems close to the orbital period minimum  in the  
range 80---90min \cite{2009MNRAS.397.2170G}. After reaching the period minimum the CVs should evolve back
toward longer periods and form a so-called bounce-back or post-period minimum systems.}
 The age of the Galaxy is old enough  for significant fraction  of the current short-period CV population to have evolved
past the orbital period minimum.
 As the objects evolve further, their luminosity is expected to decrease and their orbital periods may become larger than the period minimum and have an extremely low mass-ratio.
\begin{table}[t]
\small
    \caption{Parameters of WZ Sge and of  bounce back candidates.} 
\begin{center}
\begin{tabular}{l|lccccccccccc} \hline
NN/Object & P$_{orb}$ &    V       &  q     & M$_1$&   M$_2$ & T$_{eff}^{WD}$ & $i$ & LC$^1$   \\ 
          & (days)   &         (mag)                &         & (M$_\odot$) & (M$_\odot$) & ({\it K}) &($^o$) & & \\
  \hline
1. WZ Sge & 0.0567 & $\sim 15$ & 0.092 & 0.85 & 0.078 & 13500 & 77 & +s &  \\ 
2. GW Lib$^*$ & 0.0533 & 19.1 & 0.060 & 0.84 & 0.05 & & 11 & \\ \hline
3. V455 And$^*$ & 0.0563 & 16.5 & 0.060s &             & $>$M9 &11500 & 83 & +q \\
4. AL Com$^*$    & 0.0567 &  19.1 & 0.060          &      &    &16300&       & +q \\
5. SDSS1035 & 0.057 & 18.7 & 0.055e& 0.94 & 0.05& 10100& 83&  \\
6. SDSS1238 & 0.056 & 17.8 & 0.05   & $\sim$ 1.0 & 0.05 & 12000 & $\sim$70& +q \\
7. SDSS0804$^*$ & 0.059 & 17.8 & 0.05s  &$\sim 0.9$ & 0.045 & 13000 & $\sim 70$ & +q \\
8. EG Cnc$^*$ & 0.060 & 18.8 & 0.035s & & & 12300 & & +s\\
9. RX1050-14 & 0.062 & 17.6 & $<$0.055v& & & 13000 & $<$65& \\
10. GD552 & 0.0713 & 16.6 & $<$0.052v& & $<$0.08& 10900&$<$60 & \\
11. RE1255& 0.083& 19.0 & $<$0.064v& $>$0.9 & $<$0.08 & 12000 & $<5$ & - \\ \hline
12. SDSS1610$^{**}$ & 0.0582 & 19.0 &&&&&& +q \\
\hline
\end{tabular}
\end{center}
$^1$ light curve (LC) features: "+" LC shows  a double-hump during the orbital period; \\ "s" - during  super-outburst; "q" - during quiescence; "-"  absent of double-humps in LC. \\
$^*$ - objects which demonstrate WZ Sge-type super-outburst. \\
$^{**}$ The mass ratio of SDSS1610 is not known, however the objects shows similar observational characteristics to selected candidates and the double-humped light curve too.
\label{tabl}
\end{table}
The SDSS helped to reveal a number of  objects which can be classified as evolved beyond the period minimum and expectations are high that many more bounce-back objects can be identified among faint SDSS CVs.
In this report, we discuss  possible candidates for post-period minimum systems and  their  observational characteristics.    The  origin  of double-humped light curves is also discussed as a result of a  model of  a large cool accretion disk with spiral arms.  
\begin{figure}[b]
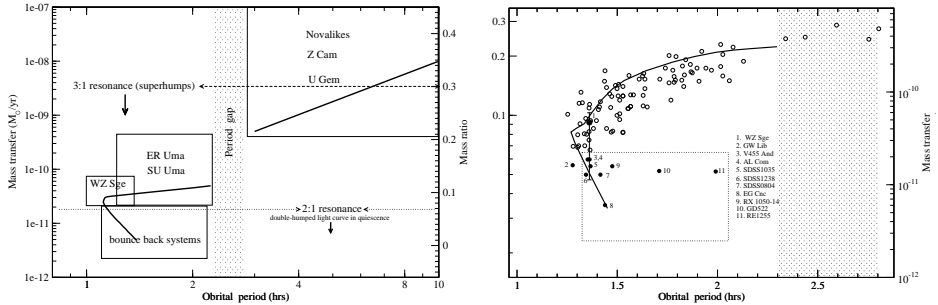

\setlength{\unitlength}{1mm}
\resizebox{10cm}{!}{
\begin{picture}(100,35)(0,0)
\put (0,0){\includegraphics[width=6.2cm, clip=]{zharfig1.eps}}
\put (65,0){\includegraphics[width=6.0cm, clip=]{zharfig3.eps}}
\end{picture}}
\caption{Left panel) The location of  bounce-back systems among other  CVs:  
Right panel) Plot of the  mass ratio and the mass transfer rate vs the orbital period. The  best bounce back candidates are enclosed in the box.}
\label{fig1}
\end{figure}
\section{The sample}

Figure\,1  illustrates the current concepts of evolution of CVs at the turning point and displays positions of bounce back system candidates on the mass-transfer rate and mass-ratio to the orbital period diagrams. Table\,1 lists the available parameters of the  post-period minimum candidates (shown within a box in Fig.1, according to  \cite{Patterson}). All systems presented in Table\,1 
show WZ Sge-type spectral characteristics in the optical range, i.e.  a mildly blue continuum with relatively weak (compared to ordinary DNe) Balmer emission lines  from the accretion disk  surrounded by broad absorptions formed in the atmosphere of  WD. Four objects (V455 And, AL Com, SDSS0804 and EG Cnc) produced WZ Sge-type super-outbursts in the near past.  Estimates of system parameters  show the presence of  relatively cool ($T_{eff} \approx 12000\pm1000$K), massive white dwarf  ($M_1\sim 0.9M_\odot$) and an extremely low value of mass ratio  $q\leq0.06$, which assumes a Jupiter-size brown dwarf as a secondary. High inclination systems frequently exhibit  light curves featuring double humps during  super-outbursts and in quiescence.  Some new phenomena were discovered in bounce-back candidates not seen anywhere else. One is the long $3.5$h spectroscopic period not related  to the  orbital  one in V455 And \cite{AAA}, secondly there are  the "brightenings"  permanently observed in SDSS1238 and occasionally seen in SDSS0804  before it went through super-outburst in March, 2006 \cite{Zhar1}. Finally, there are  "mini-outbursts"  observed in SDSS0804 \cite{Zhar2} about one year after the super-outburst. 

\begin{figure}[t]
\setlength{\unitlength}{1mm}
\resizebox{10cm}{!}{
\begin{picture}(100,45)(0,0)
\put (0,0){\includegraphics[width=5.cm, clip=]{zharfig2.eps}}
\put (55,-2){\includegraphics[width=5.8cm, bb = 130 175 543 540, clip=]{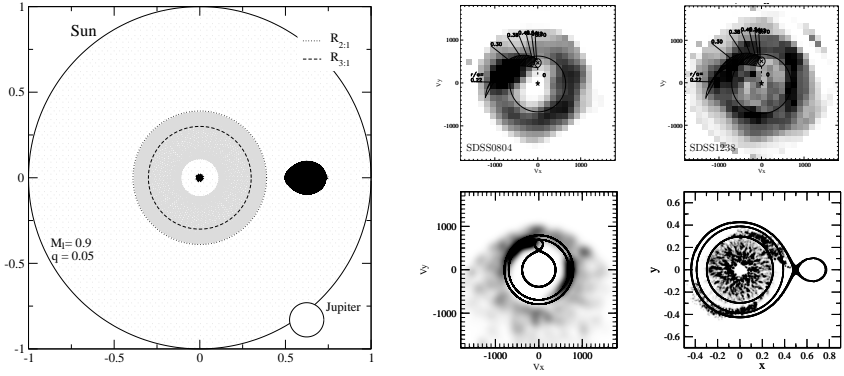}}
\end{picture}}
\caption{Left panel) The size of bounce-back system and its component in  solar radii. The 2:1 and 3:1  radiuses are shown. Right panel) The Doppler tomograms of SDSS0804 and SDSS1238 (top) and their simulation (bottom). 
}
\label{tom}
\end{figure}

\begin{figure}[t]
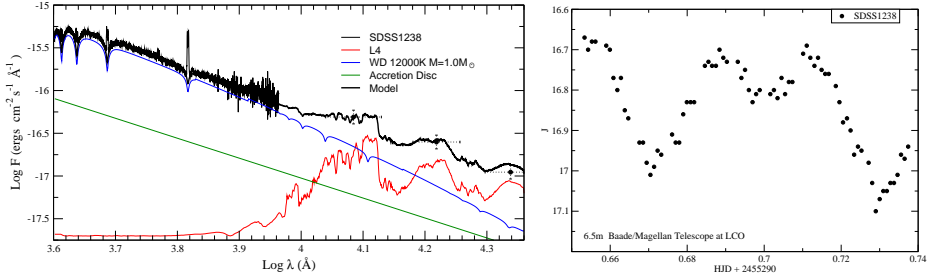

\setlength{\unitlength}{1mm}
\resizebox{10cm}{!}{
\begin{picture}(100,33)(0,0)
\put (0,0){\includegraphics[width=7.0cm, clip=]{zharfig5.eps}}
\put (72,0){\includegraphics[width=5.2cm,  clip=]{zharfig6.eps}}
\end{picture}}
\caption{Left panel) The spectral energy distribution of SDSS1238 and the result of its SED model fit\cite{Aviles}. Right panel) The J band light curve of SDSS 1238.}
\label{sed}
\end{figure}

\section{The accretion disk structure.} 
The accretion disk structure of bounce-back candidates is analyzed here based on our spectral and IR photometric 
 observations of SDSS1238 and SDSS0804 in quiescence. The conditions of the disk are inferred from Doppler tomography maps displaying internal structure on one hand,
 and on the other  from estimates of disk contribution in a system in which both stellar components are clearly visible. Doppler maps of SDSS0804 and SDSS1238 constructed using $H_\alpha$ emission line are shown in Fig.\ref{tom} 
 There is a bright spot at the expected place where the
stream of matter from the secondary collides with the accretion disk, but it overlaps with
a much larger and prolonged structure, too large to be  part of the spot. Another
extended bright region of similar size can be seen at velocity coordinates ($V_x\approx 700\ km/s$, $V_y\approx 0\ km/s$)  as well as a less bright structure at $V_x\approx -200\ km/s$, $V_y\approx -800\ km/s$. Similar
Doppler maps were obtained  during WZ Sge  super-outburst in 2001 (\cite{Baba},
\cite{Howell}) and in quiescence for the bounce-back candidate SDSS 1035\cite{Southworth} and 
were
 %
 %
%
interpreted as  evidence of
spiral waves in the disk. The formation of a spiral structure in an accretion disk of a close binary
system was predicted by Lin \& Papaloizou (1979)\cite{Lin} and explored further by various authors. 
Sawada et al. (1986) demonstrated from high resolution numerical
calculations that spirals will always form in accretion disks under tidal forces from the
secondary\cite{Sawada}. These authors actually used q = 1 in their models but observationally, such spirals
were detected in a number of systems only during outbursts of dwarf novae. Careful
examination of quiescent disks of the same systems did not reveal any spiral structures in
longer period DNe. Steeghs \& Stehle (1999) argued that little evidence of spiral arms in the
emission lines is expected in systems with low values of viscosity\cite{Steeghs}.
On the other hand, spiral arms related to 2:1 resonance can be found in systems with
extremely low mass ratio $q < 0.1$\cite{Lin}.
The bounce-back systems and others related to them, the WZ Sge stars, are examples of such objects and they are believed to have low viscosity disks.
The long outburst recurrence time in WZ Sge systems is probably explained by a very low
viscosity in their accretion disks, yet spiral arms can be observed permanently in quiescent
bounce-back systems in which, on one side there is a massive WD which gained mass during
a long accretion history, and on the side there is a late-type brown dwarf, giving  a
mass ratio of $<0.06$. WZ Sge itself shows double-humped light curve in super-outburst only. 
Fig.\,2,(bottom, right) depicts a synthetic Doppler map constructed from
a accretion disk  model that is shown on the bottom right panel.  The artificial
Doppler map reproduces the observed map in a case when there is a brightness excess within the
spiral arms. Most of the disk particles are on periodic orbits, which are most favorable
from the point-of-view of viscosity. However, the resonance dispatches some particles onto
aperiodic orbits creating viscosity perturbations, which will create heat excess.  The mechanism
is not  well established, but it is natural to assume that in these regions there will be
excess emission. The spiral arms become prominent in highly evolved systems as the result of contrast, the rest of the accretion disk
seems to contribute little to the continuum and is probably  mostly optically thin.
 \begin{figure}[t]
\setlength{\unitlength}{1mm}
\resizebox{10cm}{!}{
\begin{picture}(100,57)(0,0)

\put (5,36){\includegraphics[width=3.45cm,bb = 88 60 540 540,  angle=-90, clip=]{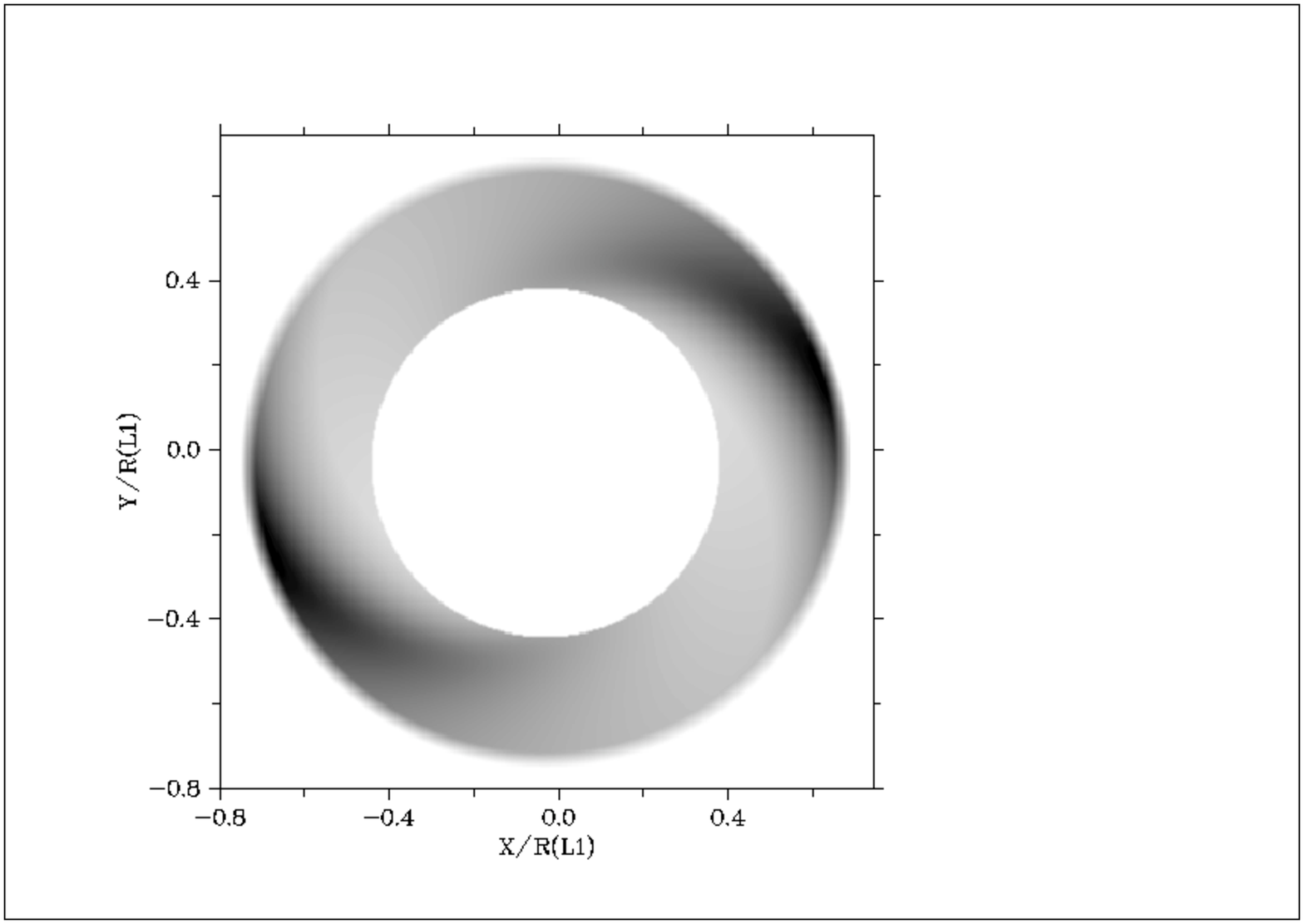}}
\put (65,36){\includegraphics[width=3.45cm,bb = 88 60 540 540, angle=-90,clip=]{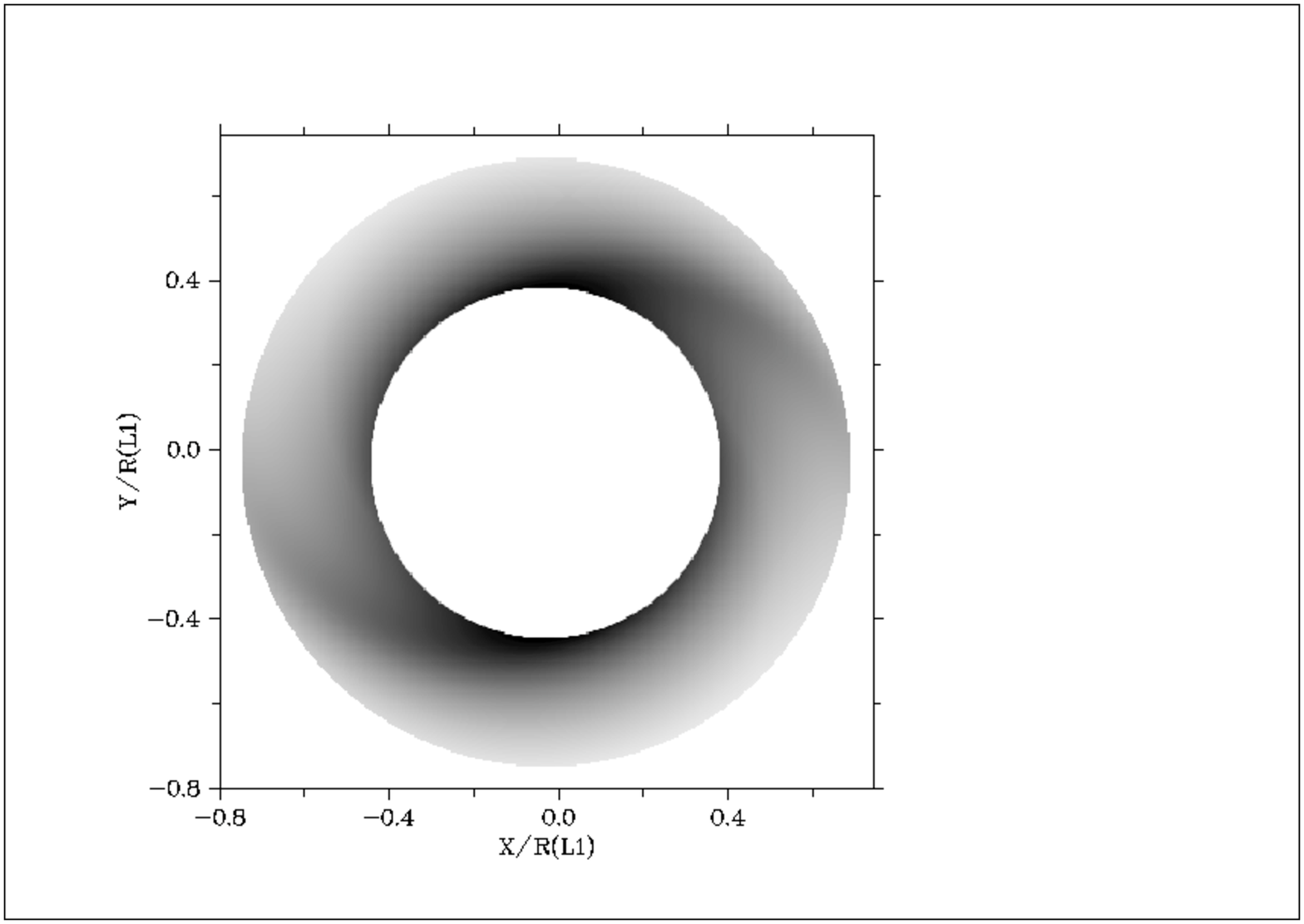}}
\put (4.5,35){\includegraphics[width=9.3cm, clip=]{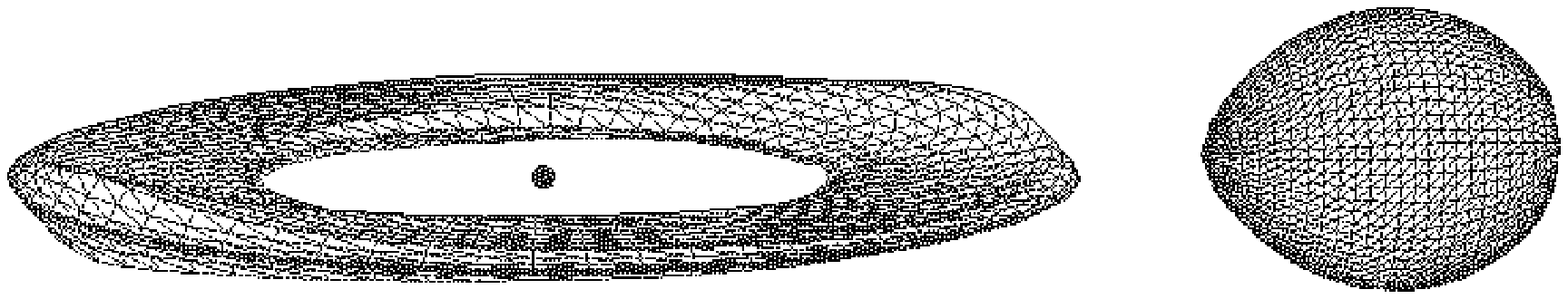}}
\end{picture}}
\caption{Top panel: Model configuration of a bounce-back system.
Bottom;  Left panel) The z-coordinate structure of the accretion disk. Right panel) The disk temperature distribution in the model.
}
\label{model}
\end{figure} 
 The visibility of both stellar components is good evidence of distinctly insignificant contribution of the accretion disk to the total optical-IR flux of the system.
 The J-band light curve  and  the optical-infrared spectral energy distribution (SED)  of SDSS1238 fitted by the three component model (WD,   a brown dwarf, and standard $F_\lambda \sim \lambda^{-7/3}$ accretion disk)\cite{Aviles}
 are shown in Fig.\ref{sed}  The model predicts very small contribution of the disk to the optical flux of  the system.
The  main uncertainty of the model is the slope of the accretion disk continuum.
The J-band light curve of SDSS 1238 corresponds to the expected   ellipsoidal shape of the secondary with a temperature gradient
 between the shaded side of the secondary and the face on WD side and also supports  a low contribution of the accretion disk in IR fluxes.
 
 The shape of the standard spectrum  ($F_\lambda \sim \lambda^{-7/3}$) of the accretion disk is based on 
 %
  the blackbody approximation of a disk element  
 intensity and $T_{eff}(r) \approx T_*(r/R_1)^{-3/4}$ relation for the radial temperature structure of  a steady state accretion disk\cite{Warner}.  
 However,  the black body accretion disk (optically thick) with a size reaching the radius of 2:1  resonance will dominate over the  radiation of the system in the optical range in hot disks or in the IR  in  low temperature  disks at practically any inclination  of the system. Therefore, the standard accretion disk model is not in agreement with the structure of accretion disks in  bounce-back systems.
 
 What do we know about the  accretion disk structure in CVs with  low accretion rates from theoretical studies? 
 Williams (1980) concludes that  accretion disks in CVs develop optically thin outer regions for mass transfer rates below about $10^{-8} M_\odot yr^{-1}$
 \cite{Williams}. Tylenda (1981) confirmed this and added that an increase in  the radius of the disk always increases the role of the thin region \cite{Tylenda}. For $\alpha <1$ the thin part of the disk is rather cool. The large part of the accretion disk is optical thin in the non-LTE regime and for $\alpha<0.1$ the temperature can drop below 5000K\cite{Dumont}. Cannizzo \& Wheeler (1984) studied  the vertical structure of a steady-state, $\alpha$-model thin accretion disk for an accreting object of $1 M_\odot$\cite{Cannizzo}. They found that for low accretion rates the disk structure  is optically thin. For $0.01< \alpha < 1$  the solution of disk equations (1)-(3) (\cite{Cannizzo}) can be double-valued with high- ($\sim 5000K$) and low- ($\sim 2000K$) temperature branches. For $\alpha> 0.1$ a warm solution is possible in the inner region of the accretion disk, but material present at larger disk radii will be in a cold state with $T < 2000K$.  Only the low temperature solution exists for $\alpha\approx 0.1$. When  $\alpha$ decreases with temperature, this tendency to develop cold solutions in quiescence in enhanced. Until now models of disk emission spectra  from such cool disks have not been calculated.  The spectrum is
  nearly flat in the range 3000-10000\AA \ for  the   model of the  cool ($T=5000K$, $\alpha=0.03$) accretion disk\cite{Idan}.

\begin{figure}[t]
\setlength{\unitlength}{1mm}
\resizebox{10cm}{!}{
\begin{picture}(100,43)(0,0)
\put (10,3){\includegraphics[width=9.5cm, clip=]{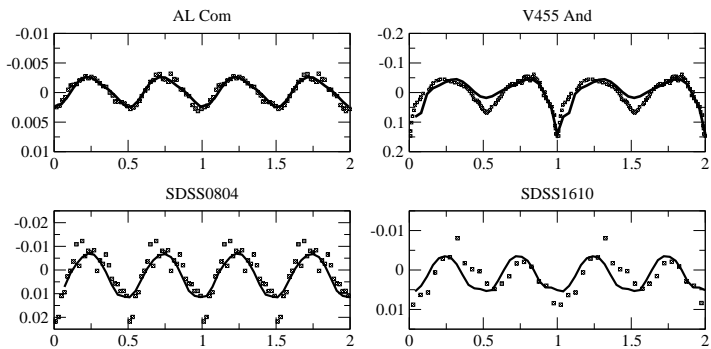}}
\end{picture}}
\caption{ The light curves of AL Com, V455 And, SDSS0804 and SDSS1610 folded with their orbital periods are shown by square points. The solid line is the light curve obtained from our model for each system.}
\label{lcmod}
\end{figure} 

	Another important aspect  of accretion disks in WZ Sge-type systems is the condition  of their inner parts. 
In the standard model, this is usually optically thick and contributes to form a continuum of the disk's spectrum.	
 However, in the case of the WZ-Sge-type system, various authors support  the idea that the inner part of the disk need to be cleared up during quiescence in order to explain the long recurrent time for super-outbursts and the behaviors of the system between super-outbursts in quiescence (see \cite{Kuulkers} and reference therein). It is not clearly understood  why the inner part of the accretion disk  is invisible. It can be caused by  evaporation\cite{Meyer}, the presence of a magnetic field of the primary WD\cite{Matthews} or an eclipsing by an optically thick spiral structure\cite{2010MNRAS.405.1397M} in high-inclination systems, or be mostly  a transparent in continuum.	 
 
 Taken into consideration  all the above
 we propose  that the accretion disks in bounce-back systems are   large (until 2:1 resonance radius), cool (about 2500K), optically thin with   hot and thick  two-spiral arm structures. There is probably a substantial  cavity in the  inner part.  

\section{The light curve simulation.}


The double-humped with orbital period light curve in quiescence  is a standing out feature of bounce-back candidates.
The light curve model\cite{Hachisu} of a CV system with a two-spiral armed thin accretion disk was adopted to bounce-back systems taking in the account positions of the bright structures in Doppler maps and a large size of the accretion disk. The small temperature gradient $T(r) \sim T(r_{in}) \times r^{-3/4}$ between the inner and outer edges of the disk and with $T\sim T(r)\times(1+\beta\times z(r)$) in spiral arms (see Fig.\ref{model}) was assumed.   
We model such discs and generate light curves which successfully simulate the observed double-humped light curves of SDSS1238, SDSS0804, SDSS1610 and V 455 And in quiescence (Fig.\ref{lcmod}).


\section{Conclusion}
We propose a likely model for a bounce-back system that explains their observed energy distributions,  Doppler tomograms and double-humped light curves at quiescence. This model consists of $\sim$12000K massive
$\sim$1M$_\odot$ white dwarf  and a late type of the brown dwarf, and a large cool ($\sim$2500K) optical thin accretion disk with
a removed/invisible/transparent inner part of the disk and two-armed hot
spirals in the outer part.


\end{document}